# Comment on "Convection in a horizontal fluid layer under an inclined temperature gradient" [Phys. Fluids 23, 084107 (2011)]

Sharifulin A.N.

Perm state technical university,614990, Perm, Russia

Ortiz-Pe´rez and Da´valos [1] investigated the stability of parallel convective flow described by the equations Eq.(9)- Eq.(11). From the article it seems that this equations obtained by the authors. Meanwhile, it is well studied theoretically and experimentally(see survey in [2]) exact solution obtained by Birikh more than half a century ago [3]:

$$v = \frac{v}{h}\frac{G}{6}(\xi - \xi^3), \quad T = Ah\left[\frac{GP}{360}(3\xi^5 - 10\xi^3 + 7\xi) - \frac{z}{h}\right].$$

Here in Birikh notation $h-$ half-width of the layer, $x\,(\xi = x/h)$ is dimensional(dimensionless) transverse coordinate, $A-$ the longitudinal temperature gradient, $G$ and $P$ are Grashof and Prandtl numbers. Introduction to the consideration of the vertical thermal gradient does not alter the velocity profile, only adds to the temperature profile the linear in $x$ term.

In the past 20 years, Birikh flow has been generalized in many papers to account for various factors of interest for applications. The generalization to the case of cylindrical geometry and the presence of rotation performed, for example, in [4,5], the effect of inclination is considered in [6]. The works, which examines the effect of vibrations can be found in the well-known monograph [7]. Among these generalizations are discussed paper too.